\documentclass[12pt]{article}
\usepackage{amsmath}%
\usepackage{amsfonts}%
\usepackage{amssymb}%
\usepackage{graphicx}
\begin{document}
\title{Simon-Gutowitz bidirectional traffic model revisited}
\author{Najem Moussa \thanks{e-mail: najemmoussa@yahoo.fr}
\\ D\'{e}partement de Math\'{e}matique et Informatique,
\\ Facult\'{e} des Sciences, B.P. 20 - 24000 - El Jadida, Morocco}
 \maketitle
\begin{abstract}
The Simon-Gutowitz bidirectional traffic model (Phys. Rev. E 57,
2441 (1998)) is revisited in this letter. We found that passing
cars get stuck with oncoming cars before returning to their home
lanes. This provokes the occurrence of wide jams on both lanes. We
have rectified the rules for lane changing. Then, the wide jams
disappear and the revisited model can describe well the realistic
bidirectional traffic.
\newline\
\end{abstract}
\newpage\
Road traffic congestion in reality is a complex phenomenon. It is
the result of interactions between many road users. The formation
of collective patterns of motion like traffic jams may be
spontaneous or induced by the presence of bottlenecks e.g., on-
and off-ramps, lane reductions, traffic lights or road works (see
the reviews [1,2]). The cellular automata (CA) models are the most
popular in the field of traffic flow modelling since they allow an
effective implementation of real-time traffic computer-simulations
(see the review [3,4]). In CA, time and space are discrete. The
space is represented as a uniform lattice of cells with finite
number of states, subject to a uniform set of rules, which drives
the behavior of the system. These rules compute the state of a
particular cell as a function of its previous state and the state
of the neighboring cells. The most popular CA model for traffic
flow on one-lane roadway is the NaSch model \cite{ns}. Despite its
simplicity, the model is capable of capturing some essential
features observed in realistic traffic like density waves or
spontaneous formation of traffic jams. Different congested traffic
states occur in other CA models: spontaneous jams caused by
velocity fluctuations, synchronous phase, wide moving jams and
stop-and-go phase \cite{Kerner}.
\newline\ A first step for describing the bidirectional traffic
is given by Lee et al \cite{Lee}. The authors have generalized the
asymmetric exclusion model. In their model no passing is allowed.
Instead oncoming traffic on the opposite lane reduces the hopping
rates of the vehicles. Simon and Gutowitz \cite{Simon} introduced
a CA model for bidirectional two-lane traffic where vehicles move
on two lanes with opposite directions. When a driver encounters a
slower forward moving vehicle, a pass will be attempted. To do
this, driver checks the density of vehicles in front that have to
be passed, i.e. the local density. If this density is low enough
the pass will be performed on condition of checking the safety
criteria on the oncoming lane.
\newline\ Up to now, Ref. \cite{Simon} is the only
paper to consider a CA model for bidirectional traffic flow. The
Simon-Gutowitz model (SG model) is a probabilistic CA which
consists of cars moving on two opposite lanes of $L$ cells with
periodic boundary conditions. Each cell is either empty, or
occupied by just one car. In the model, there exist two types of
cars: cars$^{[+]}$ moving on the lane$^{[+]}$ with positive
direction and cars$^{[-]}$ moving on the lane$^{[-]}$ with
negative direction (see figure 1a). We denote by $x$ and $v$ the
position and the speed of a vehicle at time $t$ respectively. The
maximum speed of the cars is denoted by $v_{max}$. To distinguish
between different interacting cars, several gaps and speeds are
introduced. $gap_{same}$ ($gap_{opp}$): the number of unoccupied
sites in front of a car on the same (opposite) lane.
$gap_{behind}$: the number of unoccupied sites behind the car, on
the opposite lane. $v_{same}$ ($v_{opp}$): the speed of the car
ahead on the same (opposite) lane. On the aim of making more
compact the rules of the bidirectional model, several logical
functions are introduced. \textbf{H}: true if the car is on its
home lane; \textbf{oncoming}: true if car$^{[+]}$ and cars$^{[-]}$
are face-to-face on the same lane. \textbf{Space1}: true if
($gap_{same} < l_{pass}$) AND ($gap_{opp}>l_{security}$) AND
($gap_{behind} > l_{back}$). \textbf{Space2}: true if ($gap_{opp}
> l_{security}$) AND ($gap_{behind} > l_{back}$). The parameters
$l_{pass}$, $l_{back}$ and $l_{security}$ are defined by the
following. $l_{pass}$: if $gap_{same} < l_{pass}$ AND \textbf{H}
then a pass may be attempted. $l_{back}$: the distance a driver
looks back for obstacles on the passing lane. $l_{security}$: if
$gap_{same} < l_{security}$ AND not(\textbf{H}) then the vehicle
returns immediately to its home lane. $D_{L}$: local density: the
fraction of the $l_{density} = 2v_{max}+1 sites$ in front of the
given vehicle which are occupied; $D_{limit}$: the maximum local
density for a safe pass.
\newline\ At each discrete time-step
$t\rightarrow t+1$ the system update is performed in parallel for
all cars according to the following subrules :
\newline\ \textbf{i.} \textbf{Lane changing rules}:
\newline\ 1. IF (\textbf{H} AND \textbf{Space1} AND ($D_{L}\leq D_{limit}$)
AND ($rand < p_{change}$)) THEN change lane
\newline\ 2. IF (not(\textbf{H}) AND (($gap_{same} < l_{security}$) OR
\textbf{Space2}))
THEN change lane.
\newline\ The first condition concern vehicles on their home lane that
want to change lane. When a driver encounters a slower forward
moving vehicle, a pass is attempted. However, the pass will only
be initiated if there is room far enough ahead on the passing
lane, and the number of cars in front of the vehicle it would like
to pass is small. Passing occurs randomly, even if all these
conditions are met, the probability of changing lanes is denoted
$p_{change}$. The second condition concerns vehicles in the midst
of passing. They return to their home lane if forced to by an
oncoming vehicle, or if there is space enough on the home lane
that they can return without braking.
\newline\ \textbf{ii.} \textbf{Forward moving rules}:
\newline\ 1. IF ($v \neq v_{max}$) THEN $v = v + 1$
\newline\ 2. IF((\textbf{oncoming}) AND ($gap_{same}\leq(2 × v_{max}-1))$)
THEN $v = gap_{same}/2$
\newline\ 3. IF ( (not(\textbf{oncoming})) AND ($v
> gap_{same})$) THEN $v = gap_{same}$
\newline\ 4. IF (\textbf{H} AND ($v\geq 1$) AND ($rand < p_{decel}$) AND
not(\textbf{oncoming})) THEN $v =v-1$
\newline\ 5. IF (\textbf{H} AND (\textbf{oncoming}) AND ($v
\geq 1)$ ) THEN $v = v-1$.
\newline\ If the vehicle is a car$^{[+]}$ then the vehicle moved forward
according to: $x \leftarrow x+v$. But, if the vehicle is a
car$^{[-]}$ then the vehicle moved backward according to $x
\leftarrow x-v$.
\newline\ The rule (1) reflects the tendency of drivers to drive as fast as
possible without exceeding the maximum speed limit. Rule (2)
rapidly decelerates the vehicle if there is an oncoming vehicle
too close. Rule (3) is intended to avoid collision between the
vehicles of the same type. Rule (4) randomly decelerates the
vehicle if it is on its home lane; if it is passing, it never
decelerates randomly. Finally, rule (5) breaks the symmetry
between the lanes, and thus prevents the emergence of a super jam,
i.e., a jam which may occur when each of an adjacent pair of
car$^{[+]}$ and car$^{[-]}$, one on each lane, tries to pass
simultaneously.
\newline\ In their paper, Simon and Gutowitz
compared the traffic flow of the two-lanes bidirectional traffic
with that of one-lane traffic. The improvement of the flow on one
lane compared to the one-lane model depends on the density of
vehicles on both lanes. Hence, for large densities on either both
lanes there is little difference between the one-lane traffic
model. When the density on the on-coming lane is small enough,
then the flow on the home lane can be greater than in a one-lane
model. Maximum improvement occurs near zero density on the
on-coming lane. If the density on the home lane is small then the
flow may be lower than in the corresponding one-lane model since
when oncoming cars pass other oncoming cars they can impede
traffic on the home lane.
\newline\ In this letter, we reconsider the SG model where we are interested
more especially in its congested patterns. It is clear that when
the densities on the two lanes are all very low or all very high,
the lanes will be effectively decoupled. However, in the case
where the density of one lane is low and the one of the other lane
is high, the interaction between lanes will become very important.
\newline\ In the SG model, there exist two different situations where a car
in the midst of passing may return to its home lane. The first one
is where the logical function \textbf{Space2} is satisfied. This
describes the situation when the passing car returns to its home
lane, before it is forced by an oncoming car. We find that this
condition is hardly ever satisfied if the car density on the home
lane is great enough. The second situation is when the passing car
faces an oncoming car and thus will be forced to return to the
home lane. This situation occurs if ($gap_{same} < l_{security}$).
We found that this last situation is the one the more achieved in
the SG model. Yet, almost all the passing cars provoke a head-on
collision with on-coming cars and then stop until free space
occurs on their home lane.
\newline\ In our simulations we use the following
values throughout: $l_{pass} = v$, $l_{back} = v_{max}$,
$l_{security} = 2×v_{max} +1$, $v_{max}=5$, $p_{change} = 0.5$,
and $p_{decel} =0.3$. The system size is given by $L=2000$.
\newline\ Suppose that we have a small density of cars on lane$^{[+]}$ and a
relatively high density on lane$^{[-]}$. Since free available
spaces exist on lane$^{[+]}$, cars$^{[-]}$ pass other cars$^{[-]}$
and then impede traffic on lane$^{[+]}$. Cars$^{[-]}$ in the midst
of passing will stop on lane$^{[+]}$ if they cannot return rapidly
to their home lane. Indeed, all passing cars$^{[-]}$ get stuck
with oncoming cars$^{[+]}$ before returning to their home lanes.
Consequently, the cars$^{[+]}$ regroup into wide jams. Thereafter,
passing cars$^{[-]}$ change to their home lane with keeping their
speed equal to zero. As a result, the traffic of cars$^{[-]}$ on
lane$^{[-]}$ will be delayed for a while. Obviously, the total
delayed time will be important if the total number of passing
cars$^{[-]}$ is important. Remark that, the number of passing cars
will increase if the maximum local density for a safe pass
($D_{limit}$) increases. Congested patterns in SG model are
illustrated in figure 2.
\newline\ From our daily
driving experiences in bidirectional traffic, we know that passing
cars always return to their home lane before a head-on collision
happens. That is why wide jams cannot exist in realistic
bidirectional traffic without bottleneck. To prevent the
occurrence of these wide jams it is necessary to allow returns of
cars in the midst of passing before facing an oncoming car. Hence,
we propose to rectify the lane changing rules in the SG model.
Since the occurrence of a super jam is forbidden in the SG model,
we don't need to require a free space ahead on the home lane for
the passing car, i.e., $gap_{opp}> l_{security}$. Figure 1b can
serve as an illustration. Hence, the logical function Space2
becomes:
\newline\ \textbf{Space2}: true if $gap_{behind} > l_{back}$.
\newline\ We define the revisited version of the SG model by
considering the new version of the logical function
\textbf{Space2} and by setting $D_{limit} = 1/l_{density}$.
Congested patterns in the new version of the SG model are
illustrated in figure 3. Hence, this figure shows clearly that
wide jams disappeared. On lane$^{[+]}$, cars$^{[+]}$ move freely
in spite of some obstructions caused by passing cars$^{[-]}$. On
lane$^{[-]}$, we found density waves of cars$^{[-]}$ and a small
number of passing cars$^{[+]}$ (see figure 3). Notice that, in
contrast to the old version of the SG model, passing cars$^{[-]}$
always return to their home lane with non-vanishing speed.
Besides, the head-on collisions between cars$^{[+]}$ and
cars$^{[-]}$ occur rarely in the new version of the model.
\newline\ Now we shall study the effect of
varying the densities of cars on the traffic flow in the SG model.
To do this, we fix the density of cars$^{[+]}$ and we vary the
density of cars$^{[-]}$ ($\rho^{[-]}$). Suppose that the density
of cars$^{[+]}$ is low enough. When $\rho^{[-]}$ is very large,
the two lanes are decoupled and the flow on lane$^{[+]}$ is
identical to one-lane traffic flow. With decreasing $\rho^{[-]}$,
the density of cars in front that have to be passed can be
inferior to $D_{limit}$ and then some cars$^{[-]}$ can pass on
lane$^{[+]}$. We denote by $\rho^{[-]}_{h}$ the critical density
of cars$^{[-]}$ above which the pass of cars$^{[-]}$ is forbidden.
Thereafter, these passing cars$^{[-]}$ get stuck with oncoming
cars$^{[+]}$ before returning to their home lane. This impedes
traffic of cars$^{[+]}$ on lane$^{[+]}$. From figure 4, we observe
that the flow of cars$^{[+]}$ decreases abruptly to a very small
value. Indeed, the presence of only a few passing car$^{[-]}$ can
create this abrupt decrease of the flow. With decreasing again
$\rho^{[-]}$, the flow remains constant until the density reaches
a critical density $\rho^{[-]}_{l}$. We find that $\rho^{[-]}_{l}$
is close to the critical density separating the free and congested
states in one-lane traffic model. Below $\rho^{[-]}_{l}$, the
traffic flow of cars$^{[+]}$ increases with $\rho^{[-]}$. This is
not due only to the decrease of the number of passing cars$^{[-]}$
but also to the fact that their returns to home lane become more
and more accessible.
\newline\ It is clear that $D_{limit}$ is a pertinent parameter in
the SG model. If $\rho^{[-]}$ is very large, the fraction of the
$l_{density}$ sites in front of a given car$^{[-]}$ which are
occupied is almost equal to one. Therefore, if $D_{limit}$ is
small, the probability that a car$^{[+]}$ change lanes is zero.
Hence, $\rho^{[-]}_{h}$ should decrease when one deceases
$D_{limit}$. As figure 4 shows, the new version of the SG model
presents a less important reduction of the traffic flow of
cars$^{[+]}$ than those produced by the old version of the SG
model. Furthermore, the critical density $\rho^{[-]}_{h}$ in the
new version is lower than that in the old version of the SG model.
\newline\ In figure 5 we plot the mean size of the longest
cluster of cars$^{[+]}$ as a function of $\rho^{[-]}$. For low
values of $\rho^{[-]}$, cars$^{[+]}$ regroup in clusters whose
sizes increase with $\rho^{[-]}$. When $\rho^{[-]}$ exceeds
$\rho^{[-]}_{l}$, the size of the largest cluster becomes great
and vary almost constantly with $\rho^{[-]}$. This formation of
wide jams leads to the maximal reduction of the traffic flow of
cars$^{[+]}$. If $\rho^{[-]}$ exceeds $\rho^{[-]}_{h}$, the size
of the largest cluster drops to a value equal to one. In this
case, no interaction exist between lanes and the state of
cars$^{[+]}$ will be free flow. Figure 5 illustrates clearly that
the new version of the SG model do not exhibit wide jams but only
a small clusters emerge in lane$^{[+]}$. These results are
compatible with patterns shown in figure 3.
\newline\ Suppose now that the density of cars$^{[+]}$ is relatively
high. If $\rho^{[-]}$ is low, the flow of cars$^{[+]}$ will be
greater than the flow of one-lane traffic model. As regards the
effect of $D_{limit}$, we observe that when this last increases,
the traffic of cars$^{[+]}$ is enhanced. Yet, the number of
passing cars$^{[+]}$ should increase and then will contribute
enough to the flow of cars$^{[+]}$. With increasing $\rho^{[-]}$
the flow decreases. It becomes equal to the flow of one-lane
traffic model when $\rho^{[-]}$ exceeds certain value
$\rho_{c}^{[-]}$. The results are depicted in figure 6. In the
revisited version of the SG model, the flow of cars$^{[+]}$ is
slightly superior to the flow of one-lane traffic model.
\newline\ In figure 7 we show the
variation of the mean size of the longest cluster of cars$^{[+]}$
as a function of $\rho^{[-]}$. In the old version of the SG model,
wide jams occur in lane$^{[+]}$ at low values of $\rho^{[-]}$.
These wide jams disappear when $\rho^{[-]}$ exceeds
$\rho_{c}^{[-]}$. In contrast, in the new version of the SG model,
wide jams do not exist.
\newline\ In summary, the SG model for bidirectionnel traffic flow
is revisited. If the density of cars$^{[+]}$ is small and the one
of cars$^{[-]}$ is high enough then wide jams occur in both lanes.
The occurrence of these wide jams are due principally to the fact
that almost all passing cars$^{[-]}$ get stuck with oncoming
cars$^{[+]}$ before returning to their home lanes. The traffic
flow of cars$^{[+]}$ is very small whereas the flow of
cars$^{[-]}$ is greater than in the one-lane model. We have
rectified the lane changing rules. As a result, the traffic flow
of cars$^{[+]}$ is enhanced and the wide jams disappear. We
believe that this revisited version of the SG model can describe
well the realistic bidirectional traffic.
\newpage\

\newpage\ \textbf{Figures captions}
\begin{quote}
\textbf{Figure 1}. Bidirectional model.
\newline\ \textbf{Figure 2}. Congested patterns in SG model.
On the left: lane$^{[+]}$: the black dots are cars$^{[+]}$ and the
blue dots are passing cars$^{[-]}$. On the right: lane$^{[-]}$:
the blue dots are cars$^{[-]}$ and the red dots are passing
cars$^{[-]}$ which are returned to their home lane (their speeds
are all equal to zero). The lattice size is $L=400$,
$D_{limit}=2/l_{density}$ and the lane densities are :
$\rho^{[+]}=0.05$ and $\rho^{[-]}=0.30$.
\newline\ \textbf{Figure 3}. Congested patterns in the new version
of the SG model. On the left: lane$^{[+]}$: the black dots are
cars$^{[+]}$ and the blue dots are passing cars$^{[-]}$. On the
right: lane$^{[-]}$: the blue dots are cars$^{[-]}$, the black
dots are passing cars$^{[+]}$ and the red dots are passing
cars$^{[-]}$ which are returned to their home lane (their speeds
are all different from zero. The lattice size is $L=400$ and the
lane densities are : $\rho^{[+]}=0.05$ and $\rho^{[-]}=0.30$.
\newline\ \textbf{Figure 4}. Flow of cars$^{[+]}$ as a function of
the density of cars$^{[-]}$. The density of cars$^{[+]}$ is low
enough ($\rho^{[+]}=0.05$).
\newline\ \textbf{Figure 5}. Mean size of the longuest cluster of
cars$^{[+]}$ as a function of the density of cars$^{[-]}$. The
density of cars$^{[+]}$ is low enough ($\rho^{[+]}=0.05$).
\newline\ \textbf{Figure 6}. Flow of cars$^{[+]}$ as a function of
the density of cars$^{[-]}$. The density of cars$^{[+]}$ is high
enough ($\rho^{[+]}=0.30$).
\newline\ \textbf{Figure 7}. Mean size of the longuest cluster of
cars$^{[+]}$ as a function of the density of cars$^{[-]}$. The
density of cars$^{[+]}$ is high enough ($\rho^{[+]}=0.30$).
\end{quote}

\begin{thebibliography}\
\bibitem{Helb}
{D. Helbing, Rev. Mod . Phys., \textbf{73} 1067 (2001).}
\bibitem{Mahn}
{R. Mahnkea, J. Kaupus and I. Lubashevsky, Phys. Rep. \textbf{408}
1130 (2005).}
\bibitem{chow}
 {D. Chowdhury, L. Santen and A. Schadschneider, Phys. Rep.
\textbf{329} 199 (2000).}
\bibitem{Maer}
 {S. Maerivoet and B.D. Moor, Phys. Rep. \textbf{419} 1 (2005).}
\bibitem{ns}
{K. Nagel and M. Schreckenberg, J. Phys. (France) I, \textbf{2},
2221 (1992).}
\bibitem{Kerner}
{B.S. Kerner, S.L. Klenov, D.E. Wolf, J. Phys. A: Math. Gen,
\textbf{35}, 9971 (2002).}
\bibitem{Lee}
{H.W. Lee, V. Popkov and D. Kim, J. Phys. A, \textbf{30}, 8497
(1997).}
\bibitem{Simon}
{P.M. Simon and H.A. Gutowitz, Phys. Rev.
E, \textbf{57}, 2441 (1998).}
\end{thebibliography}
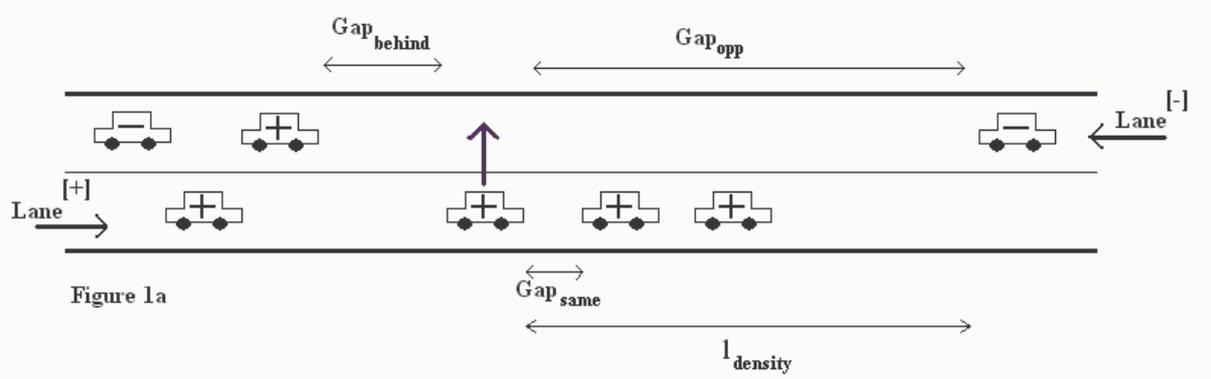
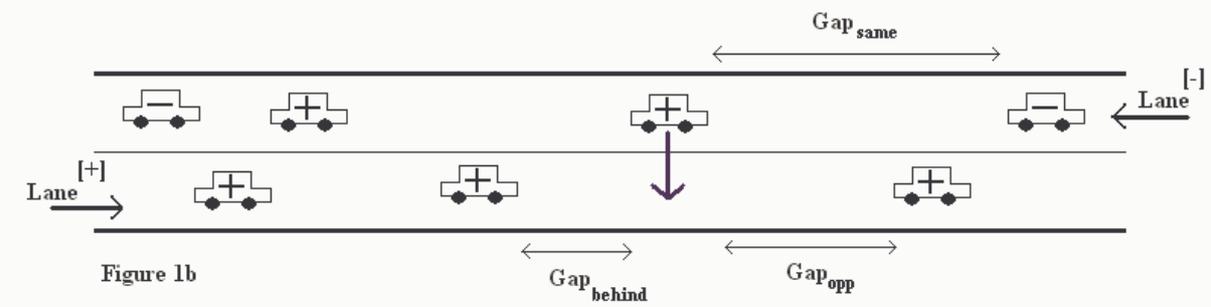
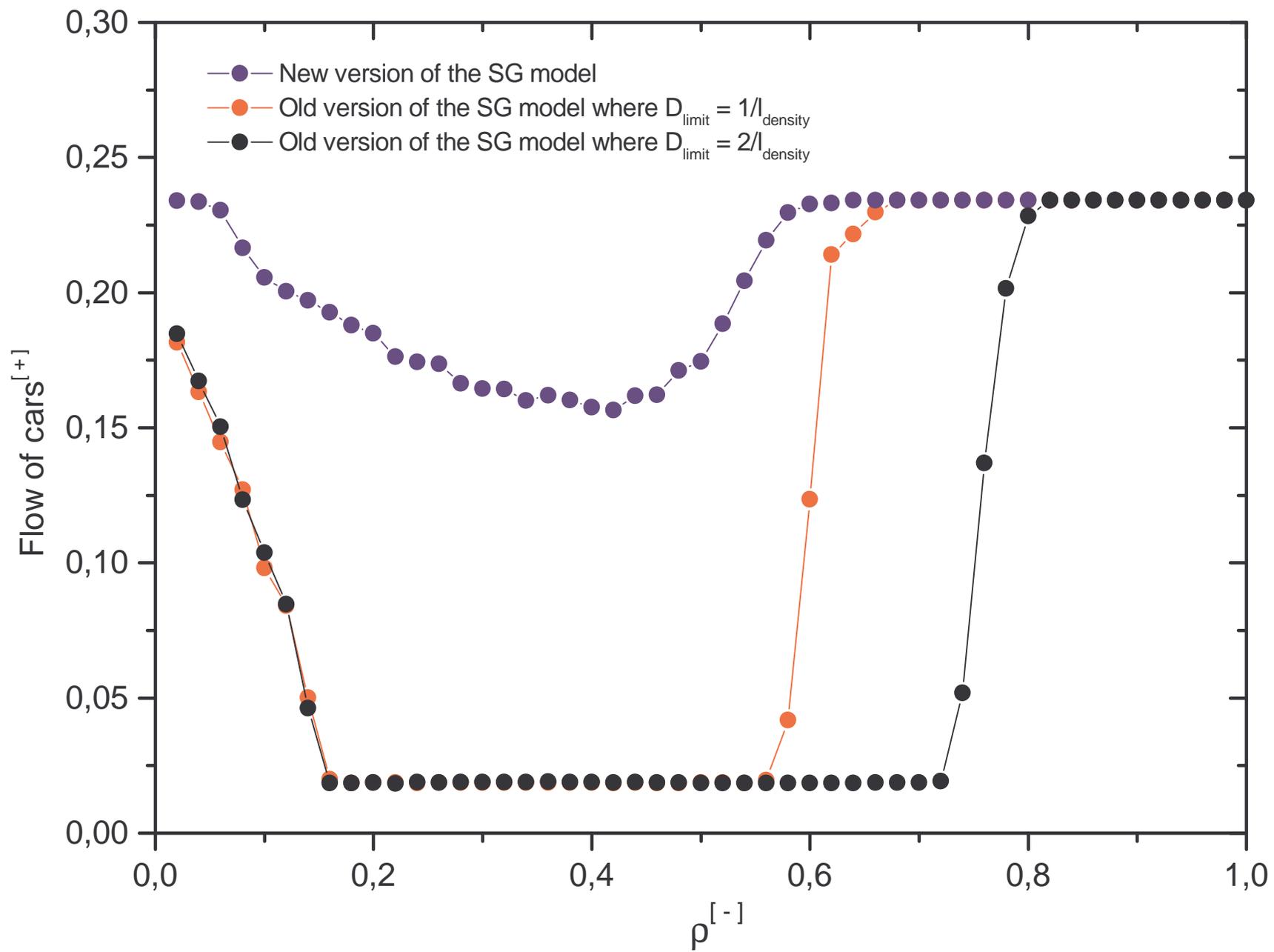
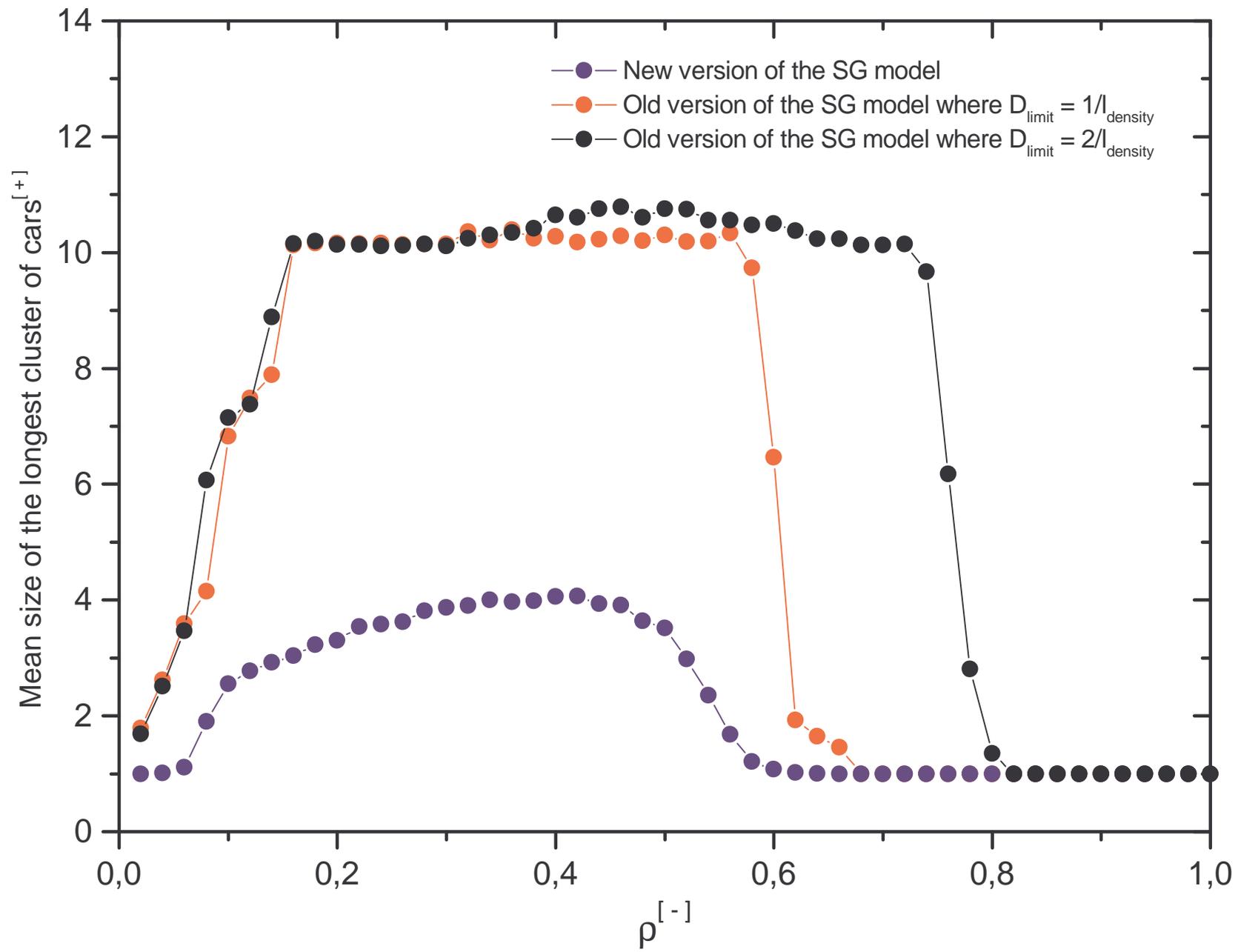
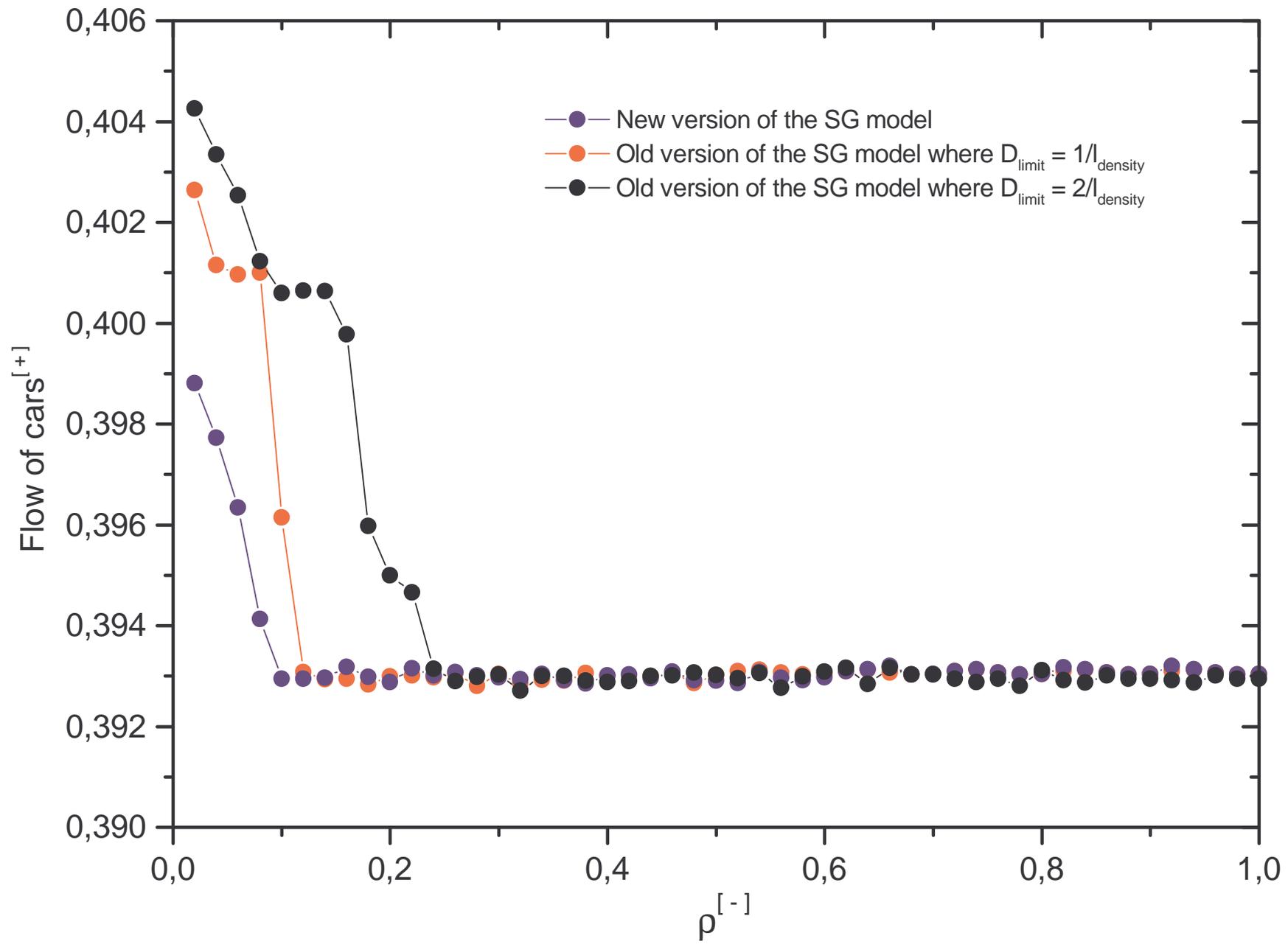
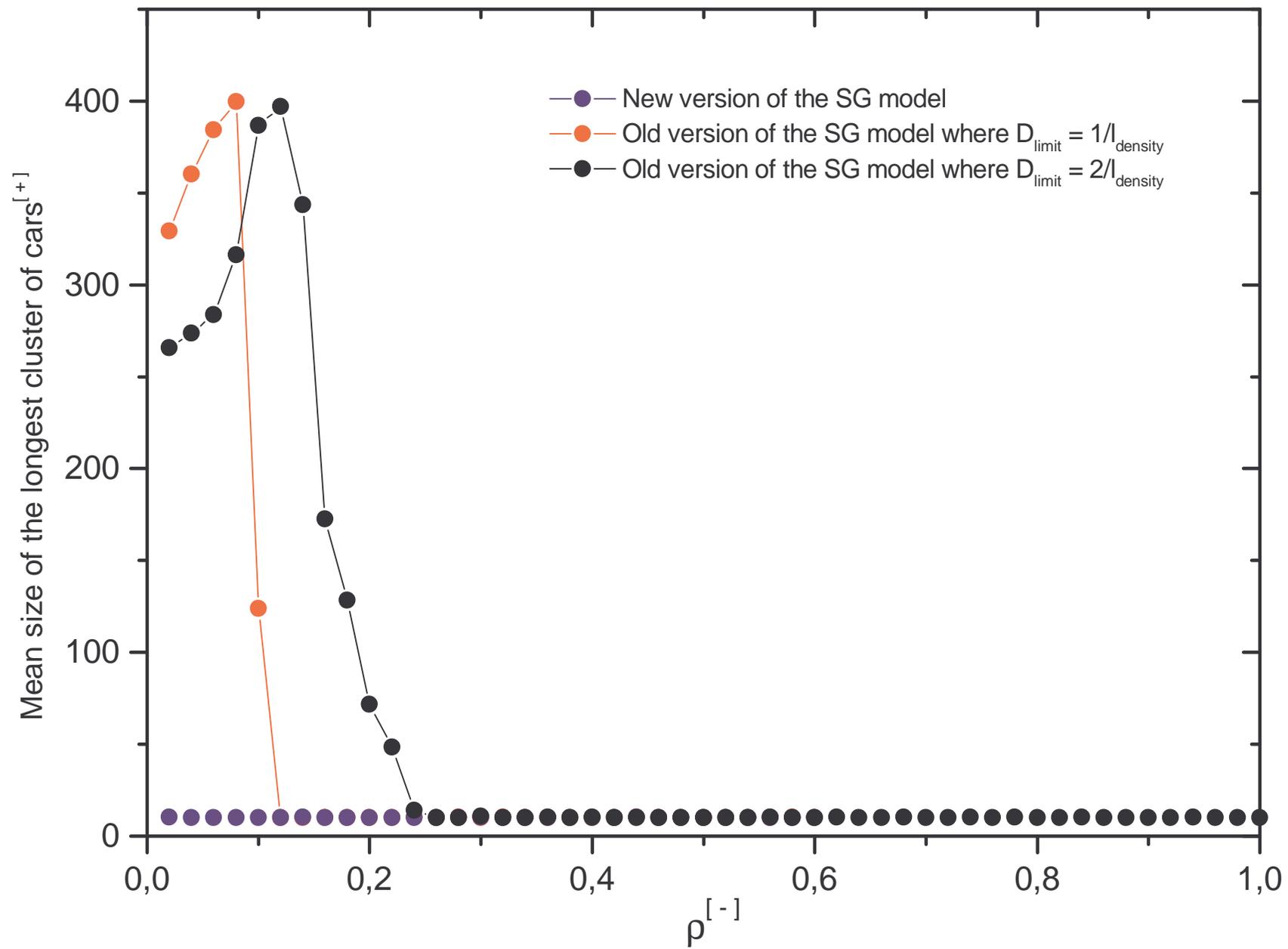
\end{document}